\documentclass[12pt]{article}
\usepackage{amsfonts}
\usepackage{latexsym}
\usepackage{amsmath,amssymb}
\usepackage{verbatim}
\usepackage{setspace}
\usepackage{cite}
\usepackage[textheight=9in, textwidth=6.5in, letterpaper]{geometry}
\def\half{{1\over 2}}
\numberwithin{equation}{section}

\def\e{{\epsilon}}

\def\l{\ell}
\def\cl{{\cal L}}

 \def\p{\partial}
 \def\bz{{\bar z}}
 
\def\0{{(0)}}
\def\1{{(1)}}
\def\2{{(2)}}
 \def\cL{{\cal L}}

\def\n{\nabla}

\def\<{\langle }
\def\>{\rangle }
\def\[{\left[}
\def\]{\right]}

\def\z{{\zeta}}

\def\sd{$\Sigma_{\rm div}$}
\newcommand{\bea}{\begin{eqnarray}}
\newcommand{\eea}{\end{eqnarray}}
\newcommand{\be}{\begin{equation}}
\newcommand{\ee}{\end{equation}}
\newcommand{\ba}{\begin{align}}
\newcommand{\ea}{\end{align}}

\renewcommand{\O}{\Omega}

\renewcommand{\epsilon}{\varepsilon}

   \makeatletter
  \let\over=\@@over \let\overwithdelims=\@@overwithdelims
  \let\atop=\@@atop \let\atopwithdelims=\@@atopwithdelims
  \let\above=\@@above \let\abovewithdelims=\@@abovewithdelims
\renewcommand\section{\@startsection {section}{1}{\z@}%
                                   {-3.5ex \@plus -1ex \@minus -.2ex}
                                   {2.3ex \@plus.2ex}%
                                   {\normalfont\large\bfseries}}

\renewcommand\subsection{\@startsection{subsection}{2}{\z@}%
                                     {-3.25ex\@plus -1ex \@minus -.2ex}%
                                     {1.5ex \@plus .2ex}%
                                     {\normalfont\bfseries}}

\begin{document}

\begin{titlepage}

\vskip4cm
\begin{center}
{~\\[140pt]{ \LARGE {\textsc{Black Hole Entropy and Soft Hair}}}\\[-20pt]}
\vskip.5cm

\vspace{0.8cm}
Sasha Haco$^{*\dagger }$, ~Stephen W. Hawking$^{*}$,~ Malcolm J. Perry$^{*\dagger  \diamond}$ and Andrew Strominger$^{\dagger}$

\vspace{1cm}

\begin{abstract}
 A set of infinitesimal  ${\rm Virasoro_{\,L}}\otimes{\rm Virasoro_{\,R}}$  diffeomorphisms are presented which  act non-trivially on the horizon of a generic Kerr black hole with spin J.  The covariant phase space formalism provides a formula for the Virasoro charges as 
 surface integrals on the horizon. Integrability and associativity of the charge algebra are shown to require the inclusion of  `Wald-Zoupas' counterterms.  A  counterterm satisfying the known consistency requirement is constructed and  yields  central charges $c_L=c_R=12J$.  Assuming the existence of a quantum Hilbert space on which these charges generate the symmetries,  as well as  the applicability  of the Cardy formula, the central charges reproduce the macroscopic area-entropy law for generic Kerr black holes. 
 \end{abstract}
\vspace{0.5cm}

\vspace{4.0cm}

\end{center}

\date{\today}

\noindent{*DAMTP, Cambridge University, Centre for Mathematical Sciences, Wilberforce Road, Cambridge CB3 0WA, UK}

\noindent{$\dagger$Center for the Fundamental Laws of Nature, Harvard University, Cambridge, MA USA}

\noindent{$\diamond$Radcliffe Institute for Advanced Study, Cambridge, MA  USA}
\end{titlepage}

\pagestyle{empty}
\pagestyle{plain}

\def\gzz{\gamma_{z\bz}}
\def\vx{{\vec x}}
\def\p{\partial}
\def\po{$\cal P_O$}
\def\cN{{\cal N}_\rho^2 }
\def\N{${\cal N}_\rho^2 ~~$}
\def\G{\Gamma}
\def\a{{\alpha}}
\def\b{{\beta}}
\def\g{\gamma}
\def\ch{{\cal H}}
\def\t{\theta }
\def\O{\Omega}
\def\cq{{\cal Q}}
\def\o{\omega}
\def\slsl{$SL(2,R)_L\times SL(2,R)_R$}
\def\vlvl{${\rm Vir_{\,L}}\otimes{\rm Vir_{\,R}}$}
\def\sb{\Sigma_{\rm bif}}
\pagenumbering{arabic}
\newpage

\begin{center}
\vspace*{\stretch{0.8}}
\setlength{\fboxrule}{1.5pt}
\fbox{\begin{minipage}{33em} \center 
{\it  We are deeply saddened to lose our  much-loved  friend and collaborator Stephen Hawking whose contributions to black hole physics remained vitally stimulating to the very end.  This paper summarizes the status of our long-term project on large diffeomorphisms, soft hair and the  quantum structure of black holes until the end of our time together.}
\end{minipage}} 
\vspace*{\stretch{1}}
\end{center}
\newpage
\tableofcontents
\section{Introduction}
Many supersymmetric or near-supersymmetric black holes in string theory admit a \vlvl\ action of nontrivial or `large' diffeomorphisms \cite{Strominger:1996sh,Horowitz:1996fn} (henceforth large diffeos) 
whose central charge can be determined by the analysis of Brown and Henneaux \cite{Brown:1986nw}.  This fact, along with a few modest assumptions, allow one to determine the microscopic entropy of the black hole and reproduce  \cite{Strominger:1997eq} the macroscopic area law \cite{Hawking:1974sw} without reliance on stringy microphysics. 

More recently, the effects of large diffeos on physically realistic black holes have been studied from a different point of view  \cite{Hawking:2016msc,Hawking:2016sgy,Flanagan:2015pxa, Averin:2016ybl, Compere:2016jwb, Sheikh-Jabbari:2016lzm, Baxter:2016nml, Compere:2016gwf, Mao:2016pwq, Averin:2016hhm, Cardoso:2016ryw, Mirbabayi:2016axw, Grumiller:2016kcp, Donnay:2016ejv, Gabai:2016kuf, Gomez:2016hxz, He:2016yhv, Tamburini:2017dig, Ammon:2017vwt, Zhang:2017geq, Bousso:2017dny, Strominger:2017aeh, Hotta:2017yzk, Mishra:2017zan, Gomez:2017ioy, Grumiller:2017otl, Chatterjee:2017zeb, Chu:2018tzu, Kirklin:2018wvq, Cvetkovic:2018dmq, Grumiller:2018scv, Chandrasekaran:2018aop, Averin:2018owq, Choi:2018oel, Donnay:2018ckb, Compere:2016hzt, Donnay:2015abr}, beginning from the observation of Bondi, Metzner, van der Burg and Sachs \cite{bms} that they can act nontrivially on the boundary of spacetime at infinity.  This paper initiates a synthesis of these approaches, and provides motivating evidence for the conjecture that the entropy of real-world Kerr black holes can be understood in a manner similar to their mathematically much better understood stringy counterparts. 
  
The large diffeos in stringy examples are not ordinarily taken to act on the entire asymptotically flat spacetime. Roughly speaking, the spacetime is divided into  two pieces. One piece contains  the black hole and the other asymptotically  flat piece has an  inner boundary surrounding a hole. The large diffeos are taken to act on the black hole. The dividing surface is often taken to be the the `outer boundary' of a decoupled near-horizon AdS$_3$ region, and the large diffeos are taken to act on this region. However, there is some ambiguity  in the choice of dividing surface, and with a suitable extension inward, the large diffeos can alternately be viewed as acting on the horizon. Indeed, when the black hole is embedded in an asymptotically flat spacetime there is no clear location to place the outer boundary of the AdS$_3$ region, and the horizon itself provides a natural dividing surface. Using the covariant phase space formalism \cite{Crnkovic:1986ex,Zuckerman:1989cx,Lee:1990nz,Brown:1992br,Iyer:1995kg,wz,Barnich:2001jy}  (see also the cogent recent review \cite{Compere:2018aar}) with a surface term reproduces the standard entropy results for BTZ black holes in AdS$_3$ from an intrinsically horizon viewpoint, albeit with a slight shift in interpretation.  Further comments on this division of the spacetime,  and the corresponding split of the Hilbert space into two pieces,  appear in the concluding section. 

Using the horizon itself as the dividing surface permits the analysis of a more general class of black holes without near-horizon decoupling regions, such as most  of those seen in the sky. It was recently shown   \cite{Hawking:2016msc,Hawking:2016sgy} that supertranslations act non-trivially on a generic black hole, changing both its classical charges and quantum state $i.e.$ generating soft hair. However, supertranslations form an abelian group and are clearly inadequate for an inference of the entropy along the lines of  the stringy analysis. As emphasized in  \cite{Hawking:2016msc,Hawking:2016sgy,Mirbabayi:2016axw,Bousso:2017dny} a richer type of soft hair, as in the stringy examples, associated to nonabelian large diffeos, is needed. 

In this paper we consider a more general class of \vlvl\ diffeos of a generic spin $J$ Kerr black hole, inspired by the discovery some years ago \cite{Castro:2010fd} of a `hidden conformal symmetry' which acts on solutions of the the wave equation in Kerr in a near-horizon region of $phase~space$ rather than $ spacetime$. In  \cite{Castro:2010fd} and subsequent  work $e.g.$ \cite{Chen:2010as,Wang:2010qv,Chen:2010xu,Becker:2010dm,Wang:2010ic,Agullo:2010hi,Chen:2010ik,Setare:2010cj,Cvetic:2010mn,Lowe:2011wu,Cvetic:2011hp,Cvetic:2012tr,Cvetic:2011dn,Compere:2012jk,Virmani:2012kw} the numerological observation  was made that, if one assumes the black hole Hilbert space  is a unitary two-dimensional CFT with $c_R=c_L=12J$, the Cardy formula reproduces the entropy. 
Here we bring this enticing numerological observation two steps closer to an actual explanation  of the entropy. First we give precise meaning to the hidden conformal symmetry in the form of an explicit set of \vlvl\ vector fields which generate it  and moreover act non-trivially on the horizon in the sense that their  boundary charges are non-vanishing. Secondly, within the covariant  formalism, we seek and find a Wald-Zoupas boundary counterterm which removes certain obstructions to the existence of a well-defined charge and gives $c_L=c_R=12J$. 

We do not herein prove uniqueness of the counterterm, attempt to tackle the difficult problem of  characterizing  `all' diffeos which act non-trivially on the black hole horizon, or show that the charges defined are integrable or actually generate the associated symmetries via Dirac brackets. These tasks are left to future investigations. For these  reasons our work might  be regarded as incremental evidence for, but certainly not a demonstration of, the  hypothesis that hidden conformal symmetry explains the leading black hole microstate degeneracy. 

Previous potentially related attempts to obtain 4D black hole entropy from a Virasoro action at the horizon  include \cite{Carlip:1999db, 
Carlip:2007qh,Guica:2008mu,Donnay:2015abr,Donnay:2016ejv,Hajian:2017mrf,Carlip:2017xne,Gonzalez:2017sfq,Setare:2018ziu}.

This paper is organized as follows. In section 2 we review prior work on hidden conformal symmetry. Section 3 presents conformal coordinates in which the Virasoro action takes the simple form presented in section 4. In sections 5 and 6 we compute the covariant right-moving Iyer-Wald Virasoro charges and identify an obstruction related to  the holographic gravitational  anomaly of \cite{Kraus:2005zm} to their associative and integrable action. A Wald-Zoupas counterterm which eliminates the obstruction is found and the central terms computed.  In section 7 we show, assuming the validity of the Cardy formula, that the  microscopic degeneracies reproduce the area law. Section 8 concludes with a general argument that all information about microstates of a generic black hole, transforming under a Virasoro generated by a large diffeo, is contained in the quantum state outside the horizon. 

Throughout this paper we use units such that $c = \hbar = k = G = 1$.

\section{Hidden conformal symmetry}
Kerr black holes with generic mass $M$ and spin $J\le M^2$ exhibit a hidden conformal symmetry which acts on low-lying soft modes \cite{Castro:2010fd}. 
The symmetry emerges, not in a near-horizon region of $spacetime$, but in the near-horizon region of $phase~ space$ defined by 
\be\label{dds} \omega (r-r_+)\ll 1,\ee
where $\omega$ is the energy of the soft mode, $r$ is the Boyer-Lindquist radial coordinate and $r_+=M+\sqrt{M^2-a^2}$, with $a={J \over M}$, is the location of the outer horizon. This simply states that the soft mode wavelength is large compared to the black hole.  One way to see the emergent symmetry is from the fact that the explicit near-horizon wave functions of soft modes are hypergeometric functions of $r$, and therefore  fall into representations of $SL(2,R)$.
In fact, the scalar wave equation for angular momentum $\ell$ can be written in this region \cite{Castro:2010fd} as the formula for the Casimir either of an $SL(2,R)_L$ or an $SL(2,R)_R$, with conformal weights 
\be (h_L,h_R)=(\ell,\ell).\ee
A suitably modified formula applies to spinning fields.  Another signal of the symmetry is that the near region contribution to the soft absorption cross sections can be written\footnote{See \cite{Castro:2010fd} for  a derivation and discussion of the range of validity of this expression.} 
\be\label{abs} {\cal P}_{abs}\sim T_L^{2h_L-1}T_R^{2h_R-1}\sinh({\o_L \over 2T_L}+{\o_R \over 2T_R} )\big|\G(h_L+i{\o_L \over 2\pi T_L})\big|^2  \big|\G(h_R+i{\o_R \over 2\pi T_R})\big|^2.\ee
Here the left and right temperatures are defined by
\be \label{tmp} T_L={r_++r_- \over 4\pi a}, ~~~ T_R={r_+-r_-\over 4\pi a},\ee
with $r_-=M-\sqrt{M^2-a^2}$ and the left and right soft mode energies are 
\be\label{ewt} \o_L={2M^2 \over a}\o,~~\o_R={2M^2 \over a}\o-m,\ee
with $(\o,m)$ the soft mode energy and axial component of angular momentum. 
The left/right temperatures and entropies are thermodynamically conjugate, as follows from\be \delta S_{BH}={ \o_L\over T_L}+{\o_R \over T_R},\ee 
where $S_{BH}=2\pi Mr_+$ is the Kerr black hole entropy. 

Equation \eqref{abs} is precisely the well known formula for the absorption cross section of an energy $(\o_L,\o_R)$ excitation of a 2D CFT at temperatures $(T_L,T_R)$.  This motivates  the 
hypothesis  that the black hole is itself a thermal 2D CFT and transforms under a \vlvl\ action. Motivated by this, in the spirit of \cite{Hawking:2016msc,Hawking:2016sgy},  in section 4 below we explicitly realize the hidden conformal symmetry in the form of  \vlvl\  diffeos which  act non-trivially  on the black hole horizon.\footnote{We wish to note however that there may also exist, as in the Kerr/CFT \cite{Guica:2008mu} context, an alternate holographic formulation with a left Virasoro-Kac-Moody symmetry, where the Kac-Moody zero mode generates right-moving translations   \cite{Detournay:2012pc}, which surprisingly in some cases provides an alternate explanation for example of  formulae like  \eqref{abs}. Indeed with the exciting   recent progress in understanding the underlying  warped conformal field theories \cite{Guica:2017lia,Aharony:2018ics,Bzowski:2018pcy} this latter possibility is looking the more plausible for the case of Kerr/CFT.   Investigation  of hidden Virasoro-Kac-Moody  symmetries for generic black holes is left to future work.  }  
We begin by recalling  the coordinate transformation \cite{Castro:2010fd} which most clearly exhibits the conformal structure. 

\section {Conformal coordinates} The Kerr metric in Boyer-Lindquist coordinates is 
\bea ds^2&=&-\big(1-{2Mr \over \rho^2}\big)dt^2+\big(r^2+a^2+{2a^2Mr \sin^2\theta \over \rho^2}\big)\sin^2\theta d\phi^2-{4aMr \sin^2\theta \over \rho^2}d\phi dt\cr &&~~~~~~~~~~~~~
 +{\rho^2 \over \Delta}dr^2+\rho^2 d\theta^2,\eea
where \bea
~~~~~~\rho^2=r^2+a^2\cos^2\theta, ~~~~~~\Delta=r^2+a^2-2Mr.\eea
Conformal coordinates  are \cite{Castro:2010fd}\footnote{ The inverse transformation is \bea \phi&=&{1 \over 4\pi T_R}\ln{w^+(w^+w^-+y^2) \over w^-},\cr  r&=&r_++4\pi a T_R{w^+w^-\over y^2}, \cr t&=&{M(T_R+T_L) \over T_R}\ln{w^+\over w^-}+{M(T_L-T_R)\over T_R}\ln (w^+w^-+y^2).
\eea}
\bea
 w^+&=& \sqrt{r-r_+ \over r-r_-}e^{{2\pi T_R}\phi},\cr
 w^-&=& \sqrt{r-r_+ \over r-r_-}e^{{2\pi T_L}\phi-{t \over 2M}},\cr
y&=& \sqrt{r_+ -r_-\over r-r_-}e^{{\pi(T_R+T_L)}\phi-{t \over 4M}}.\eea The past horizon is at $w^+=0$, the future horizon at $w^-=0$ and the bifurcation surface $\sb$ at $w^\pm=0$. Under azimuthal identification $\phi \to \phi+2\pi$ one finds
\be\label{idn} w^+ \sim e^{4\pi^2 T_R}w^+, ~~ w^- \sim e^{4\pi^2 T_L}w^-, ~~
y\sim e^{2\pi^2 (T_R+T_L)}y.\ee
This is the same as the identification which turns $AdS_3$ in Poincar\'e coordinates into BTZ with temperatures $(T_L,T_R)$ where the $w^\pm$ plane becomes thermal Rindler space  \cite{Maldacena:1998bw}. It is for this reason that conformal coordinates are well-adapted to an analysis of 4D black holes mirroring that of the 3D BTZ black holes. 
  To leading and subleading  order around the bifurcation surface, the metric becomes
  \bea \begin{split} ds^2&= {4 \rho_+^2 \over y^2} d w^+ dw^-
+  {16 J^2 \sin^2\theta \over y^2 \rho_+^2} dy^2 +\rho_+^2 d\theta^2 \\[6pt]  &
- {2w^+ (8\pi J)^2 T_R(T_R+T_L)  \over y^3 \rho_+^2} dw^- dy \\[6pt]  &
+ {8 w^- \over y^3 \rho_+^2} \big(- (4\pi J)^2T_L(T_R+T_L) + (4 J^2 + 4\pi J a^2 (T_R+T_L)  + a^2 \rho_+^2) \sin^2\theta \big) dw^+ dy \\[6pt] &
+ \cdots, \end{split} \eea  
where corrections are at least second order in $(w^+,w^-)$. The volume element is \be \e_{\theta y + -}={8J\sin \t \rho_+^2 \over y^3}+\cdots .\ee 

\section{Conformal vector fields}
Consider the vector fields 
 \be \zeta(\e)= \e \p_++\half\p_+\e y\p_y,\ee
 where $\e$ is any function of $w^+$. 
 These obey the Lie bracket algebra
 \be \[ \z(\e),\z(\tilde \e )\]=\z(\e\p_+\tilde \e-\tilde \e\p_+ \e).\ee
 We wish to restrict $\e$ so that $\z$ is invariant under $2\pi$ azimuthal rotations \eqref{idn}. A complete set of such functions is\footnote{  \eqref{csf}, \eqref{dsg} are the same  restrictions encountered in the quotient of planar AdS$_3$ to BTZ, or 2D Minkowski to thermal Rindler \cite{Maldacena:1998bw}. They imply that the $\z_n$ ($\bar \z_n$) are periodic in imaginary right (left) `Rindler time' $2\pi\ln w^+$ ($2\pi \ln w^-$) with period $2\pi T_R$ ($2\pi T_L$) as in \eqref{csf}. }
 \be\label{csf}\e_n={2 \pi T_R}(w^+)^{1+{in \over 2 \pi T_R}}.\ee
The corresponding vector fields $\z_n\equiv \z(\e_n)$ obey the centreless Vir$_R$ algebra
 \be [\z_m,\z_n]=i(n-m)\z_{n+m}.\ee
 The zero mode is 
 \be\label{rtj} \zeta_0=2\pi T_R( w^+\p_++\half y\p_y )=\p_\phi+{2M^2 \over a} \p_t=-i\omega_R,\ee
 where the right moving energy $\o_R$ is defined in \eqref{ewt}.

 Similarly,
  \bea\label{dsg} \bar \zeta_n&=&\bar \e_n \p_-+\half\p_-\bar \e_ny\p_y,\cr
\bar\e_n&=&{2 \pi T_L}(w^-)^{1+{in \over 2 \pi T_L}},\eea
  with 
  \be \bar \z_0=-{2M^2 \over a}\p_t=i\o_L \ee
  obey the centreless Vir$_L$ algebra
 \be [\bar \z_m,\bar \z_n]=i(n-m)\bar \z_{n+m},\ee
and the two sets of vector fields commute with one another 
 \be [\z_m,\bar \z_n]=0.\ee Note that the \vlvl\ action maps the `$\theta$-leaves' of fixed polar angle to themselves. $\z$ preserves the future horizon and $\bar \z$ the past horizon. 
 
The Frolov-Thorne vacuum density matrix for a Kerr black hole is (up to normalization) 
\be \rho_{FT}=e^{-{\o \over T_H}+{\Omega m \over T_H}},\ee
where $T_H={r_+-r_-\over 8\pi Mr_+}$ and $\Omega={a \over 2Mr_+}$ are the Hawking temperature and angular velocity of the horizon, with $\o$ and $m$ being interpreted here as energy and angular momentum operators. Rewriting this in terms of the eigenvalues of the zero modes $\z_0$ and $\bar\z_0$ one finds simply  
\be \label{rst}  \rho_{FT}=e^{-{\o_R \over T_R}-{\o_L \over T_L}}.\ee
This is a restatement of the fact that  $\o_{R,L}$ is thermodynamically conjugate to $T_{R,L}$.

For future reference the only non-zero  covariant  derivatives of $\z$ on the bifurcation surface $\sb$ are 
 \bea 
  \n_+\z^+=-\G^-_{y-}\z^y,~~
 \n_-\z^-=\G^-_{y-}\z^y,~~
  \n_+\z^y=\p_+\z^y, ~~
    \n_\theta\z^y=\G^y_{\theta y}\z^y, ~~
 \n_y\z^\theta=\G^\theta_{yy}\z^y, \eea
while the only non-zero metric deviations on the bifurcation surface are 
 \bea \label{tri}
 \cl_\z g_{y+}=g_{yy}\p_+\z^y,~~~\cl_\z g_{+-}=g_{y-}\p_+\z^y.\eea
Similar formulae apply to $\bar \z$.
 
\section{Covariant charges}
In this section we construct the linearized covariant charges $\delta \cq_n\equiv \delta \cq(\z_n,h;g)$ associated to the diffeos 
$\z_n$ acting on the horizon. The construction of covariant charges has a long history including \cite{Crnkovic:1986ex,Zuckerman:1989cx,Lee:1990nz,Brown:1992br,Iyer:1995kg,wz,Barnich:2001jy}. Formally, the linearized charges  are expected to  generate the linearized action, via Dirac brackets, 
of $\z_n$ on the on-shell linearized fluctuation $h$ around a fixed background $g$. The formal argument proceeds from the fact that they   reduce to the covariant symplectic form with one argument the $\z$-transformed perturbation $h$.  However, in practice many subtleties arise when attempting to verify such expectations. Among other things one must reduce, via gauge fixing and the application of the constraints, with careful analyses of zero modes and boundary conditions,  to a physical phase space on which the symplectic form is nondegenerate.  Various obstructions may arise, such as non-integrability of the charges or violations of associativity which necessitate the addition of boundary counterterms as discussed for example in \cite{Lee:1990nz,Iyer:1995kg,wz,Barnich:2001jy,Chandrasekaran:2018aop}. 

 In the much simpler case of horizon supertranslations of Schwarzschild, it was verified in full detail \cite{Hawking:2016sgy}
 that the linearized charges $\delta \cq_f$ do indeed generate the linearized symmetries as expected. Moreover, the  $\delta \cq_f$ were in this case recently explicitly integrated to the full horizon supertranslation charges $\cq_f$ \cite{Chandrasekaran:2018aop}.
 The $\delta \cq_n$ of interest here are significantly more complicated than their supertranslation counterparts $\delta \cq_f$.  We leave  a comprehensive analysis of $\delta \cq_n$  in the style of \cite{Hawking:2016sgy} to future work, and the present  analysis should therefore be regarded as a preliminary first step.  

The construction of covariant charges has been reviewed in many places ($e.g.$ \cite{Compere:2018aar}) and is recapped in the appendix. The general form for the linearized charge associated to a diffeo $\z$ on a surface $\Sigma$ with boundary $\p\Sigma$  is \cite{wz} \be \label{charge}\delta \cq=\delta \cq_{IW}  +\delta \cq_X.\ee
Here the Iyer-Wald charge is 
\be \delta \cq_{IW}(\z,h;g)={1 \over 16\pi }\int_{\p\Sigma}*F_{IW},\ee
with $F_{IWab}$ explicitly  given by
\be \begin{split} F_{IWab} = \frac{1}{2}\n_a\zeta_bh
+\n_ah^c{}_b \zeta_c
+\n_c\zeta_a\ h^c{}_b 
+\n_ch^c{}_a\ \zeta_b
-\n_ah\ \zeta_b - a \leftrightarrow b, \end{split} \ee where the variation $h^{ab}$ is defined by $g^{ab} \to g^{ab} + h^{ab} $ and $h = h^{ab} g_{ab}$.
The Wald-Zoupas counterterm is 
\be \delta \cq_X={1 \over 16\pi }\int_{\p\Sigma}\iota_\z (* X),\ee
where $X$  is a spacetime one-form constructed from the geometry and linear in $h$.\footnote{$*X$ is often denoted $\Theta$.} $X$ is not a priori fully determined by the considerations of \cite{Iyer:1995kg,wz}, where its precise form is left as an ambiguity. Ultimately one hopes it is fixed by consistency conditions such as integrability and the demand that the charges generate the symmetry via a Dirac bracket as in \cite{Hawking:2016sgy}, or in the quantum form by action on a Hilbert space. In practice the determination of $X$ has been made on a case-by-case basis. Our case involves a surface $\Sigma$  with interior boundary on the far past of  the future horizon, namely the bifurcation surface $\sb$ at $w^\pm=0$.  The boundary charge on $\p\Sigma=\sb$ is the black hole contribution to the charge. 
We will find below consistency conditions that require a nonzero $X$. A candidate that enables them to be satisfied is  simply
\be \label{counter}  X = 2dx^ah_a^{~b} \Omega_b, \ee where $\Omega_a$ is the H\'a\'{\j}i\v cek one-form,
\be \Omega_a = q_a^c n^b \n_c l_b, \ee 
a measure of the rotational velocity of the horizon. 
Here the null  vectors $\l^a$ and $n^a$  are both normal  to $\sb $ and normalized such that $\l \cdot n=-1$. $\l$ ($n$) is taken to be normal to the future (past) horizon.\footnote{ $\l$ and $n$ must be invariant under $2\pi$ rotations which act in conformal coordinates as \eqref{idn}. This is satisfied by  $\l \sim y^{{2 T_R \over T_R + T_L}} \p_+ , n \sim y^{{2 T_L \over T_R + T_L}} \p_- $ on $\sb$.  These conditions  uniquely fixes $\l$ and $n$ up to a smooth  rescaling under which $X_a\to \p_a\phi$. We could  fix this ambiguity by demanding $e.g.$ that $\Omega$ be divergence-free on $\sb$ but this condition will not be relevant at the order to which  we work.}
$q_{ab}=g_{ab}+\l_an_b+n_a\l_b$ is the induced metric on $\sb$.\footnote{ See for example\cite{Gourgoulhon:2008pu}
for a nice review of hypersurface geometry in the context of black holes. }

 As a check on the normalization, we note that 
 \be\label{nm} \delta \cq(\p_t,\delta_Mg;g)=1.\ee
Here $\delta_Mg$ is the linearized variation of the Kerr metric at fixed $J$. The Wald-Zoupas term $\delta \cq_X$  does not contribute to this computation. 

We are especially interested in the central term in the Virasoro charge algebra. When the charge is integrable $and$ there is a well-defined (invertible and associative) Dirac bracket $\{ ,\} $ on the reduced phase space, or in quantum language when $\cq_m$ is realized as an operator generating the diffeo $\z_n$ on a Hilbert space, one has  
\be \{  \cq_n, \cq_m\}=(m-n)\cq_{m+n}+K_{m,n}, \ee
where the central term is given by
\be \label{cterm} K_{m,n}=\delta \cq(\z_n,\cL_{\z_m}g;g).\ee
Moreover, under these conditions, it has been proven (as reviewed in \cite{Compere:2018aar})  that the central term must be constant on the phase space and given, for some constant $c_R$  by 
\be K_{m,n}={c_R m^3\over 12}\delta_{m+n},\ee
up to terms which can be set to zero by shifting the charges.

 In order to evaluate the charge and the central terms we must specify falloffs for $h^{ab}$ near $\p\Sigma=\sb$.  One might demand that all components of  $h^{ab}$ (which is always required
to be on shell) approach finite functions at $\sb$ at some rate as in \cite{Chandrasekaran:2018aop}. However this condition is violated by the $h^{ab}$ produced  by the large diffeos $\z_n$. We accordingly augment the phase space to allow for these pure gauge modes as well as the on-shell non-gauge modes that approach finite values at $\sb$\footnote{The details of these rates are important for a complete investigation of integrability. We also restrict here to the phase space of fixed $J$. This is an analog of  fixing the number of branes in string theory, which indeed  in some cases is $U$-dual to the higher-dimensional angular momentum.}.  These oscillate periodically in the affine time along the null generators and do not approach a definite value at $\sb$, which is at infinite affine distance from any finite point on the horizon. Were they not pure gauge, such oscillating perturbations would have infinite energy flux and would be physically excluded. In the (non-affine) null coordinate $w^+$ along the horizon these modes can have poles at $w^+=0$. We will find that the charges are nevertheless well-defined and have a smooth $w^+\to 0$ limit  with such pure gauge excitations. Moreover, the emergence of a nonvanishing central term relies on the poles: since $\z$ is actually tangent to $\sb$ precisely at $w^+=0$,  the $\delta\cq_X$  vanishes unless the perturbation produces a $w^+$-pole in $X$.\footnote{In \cite{Chandrasekaran:2018aop} it was shown that central terms cannot appear in the absence of poles.} We will define and compute these counterterms by working at small $w^+$ and then taking the limit. This amounts to approaching $\sb$ along the future horizon.

To evaluate the central term we take $\z=\z_m$ and $h^{ab}=\cL_{\z_n}g^{ab}$. 
It turns out that nonzero contributions to $K_{n,m}$ from  $\delta \cq_{IW}$ come only from the component $F_{IW}^{-y}$ in the form\be {1 \over 16 \pi }\int_{\sb}d\theta dw^+\e_{\theta+-y}F_{IW}^{-y} .\ee  The range of $w^+\sim e^{4\pi^2T_R}w^+$ goes to zero as $\sb$ is approached, so this expression naively vanishes. However, using the relation
  \be \lim_{w^+_0\to 0}
 \int_{w^+_0}^{w^+_0e^{4 \pi^2 T_R}} {dw^+ \over w^+} = 4 \pi^2 T_R, \ee such terms can nevertheless contribute as $\p_+\z^y$ and $h^{-y}$ develop $1 \over w^+$ poles 
for $w^+ \to 0 $.
One finds, after some algebra, \be  \label{tip}F^{-y}_{IW}  =-4 h_m^{y-} \z_n^y \G^-_{y-},\ee
where 
\be h_m^{-y}=g^{+-}\p_+\z^y_m\ee
has the requisite  pole in $w^+$. 
Integrating over the sphere gives 
\be\label{iwf} K_{IW n,m}=2J{T_R \over T_L+T_R}m^3 \delta_{n+m}.\ee

Temperature dependence of the central term \eqref{iwf}  violates the theorem \cite{Compere:2018aar} that it must be constant on the phase space. Hence there is an obstruction to constructing and integrating the charges $\delta\cq_{IW}$ with well-defined associative Dirac brackets, the existence of which is assumed in the theorem. We seek to remove this obstruction on the phase space of constant $J$ by a suitable choice of $X$. However, we wish to stress the absence of this obstruction is necessary, but not a priori sufficient, for  $\delta \cq$ to exist as an operator on a Hilbert space with all the desired properties including integrability. This is left to future investigations. Moreover, we have not shown that \eqref{counter}  is unique in eliminating this obstruction. 

The obstruction is eliminated by including the Wald-Zoupas contribution $K_{Xm,n}=\delta \cq_X(\z_n,\cL_{\z_m}g;g)$, which after integration over $\sb$ gives  \be \label{xf}K_{X n,m}=J{T_L - T_R \over T_L+T_R}m^3 \delta_{n+m}.\ee
Adding terms \eqref{iwf} and \eqref{xf} then yields the central charge 
\be c_R=12J.\ee

\section{Left movers}
In order to compute the left-moving charges on $\sb$, it is necessary to evaluate \eqref{charge} with $\z = \bar{\z}_m$ and $\bar{h}^{ab} = \mathcal{L}_{\bar{\z}_n} g^{ab}$. Now the relevant contribution to $\bar{K}_{m,n}$ comes only from $F_{IW}^{+y}$. On the past horizon, the range of $w^- \sim e^{4\pi^2 T_L} w^-$ now goes to zero as $\sb$ is approached but again one finds the appearance of poles for $w^-\to0$, coming from terms such as $\p_- \bar{\z}^y$ and $\bar{h}^{+y}$. $F_{IW}^{+y}$ can be evaluated to be,
\be F_{IW}^{+y} = - 4 \bar{h}^{y+}_m \bar{\z}_n \G^+_{+y} , \ee where \be \bar{h}^{+y}_m = g^{+-} \p_-\bar{\z}^y_m \ee has a pole in $w^-$. Integrating over the sphere gives 
\be \bar{K}_{IW n,m} = 2 J {T_L \over T_R + T_L} m^3 \delta_{m+n}. \ee 

Since $\sb$ is being approached from  the past horizon, the vector fields $\l^a$ and $n^a$ are now defined so that $\l$ is normal to the past horizon and $n$ is normal to the future horizon. Again, both are null and satisfy $\l \cdot n = -1$. An analysis of the periodicities gives
\be \l \sim y^{{2 T_L \over T_R + T_L}} \p_-, \; n \sim y^{{2 T_R \over T_R + T_L}} \p_+. \ee 
The resulting term involving $X$ integrates to 
\be \bar{K}_{X n,m} =  J {T_R - T_L \over T_R + T_L}m^3 \delta_{m+n}.  \ee 
The sum of these two terms yields 
\be c_L = 12J . \ee 

We note that the Wald-Zoupas counterterm $\delta \cq_X$ contributes only to $c_L-c_R$ and not $c_L+c_R$ and hence may be related to the holographic gravitational anomalies discussed in \cite{Kraus:2005zm}.

   \section{The area law}
Using  $c_L=c_R=12J$ as given above, the temperature formulae \eqref{tmp} and the Cardy formula
\be S_{Cardy}={\pi^2 \over 3}(c_LT_L+c_RT_R),\ee
yields the Hawking-Bekenstein  area-entropy law for generic Kerr
\be S_{BH}=S_{Cardy}=2\pi Mr_+= {Area \over 4 }.\ee

\section{Discussion}
In this concluding section we give a formal argument that, whenever black hole microstates are in representations of  large-diffeomorphism-generated  Virasoro algebras, as conjectured  for Kerr in this paper, the black hole Hilbert space must be contained within 
the Hilbert space of states outside the black hole. The observations apply equally  to the case discussed here and to the stringy black holes with near-AdS$_3$ regions. Our argument is a refined and sharpened version of those made elsewhere from different perspectives and is perhaps in the general spirit, if not the letter, of black hole complementarity.\footnote{  See \cite{Harlow:2014yka} for a recent review. }

Consider a hypersurface \sd\ which divides the black hole spacetime into a black hole region and an asymptotically 
flat region with a hole. \sd\ may be taken to be  the stretched horizon, the event horizon or in stringy cases   the outer boundary of an AdS region: for the purposes of microstate counting the difference will be subleading and the distinction irrelevant.  For a scalar field theory on such a fixed geometry it is reasonably well understood how to 
decompose the full Hilbert space $\ch_{\rm full}$ of scalar excitations on a complete spacelike slice which  goes through the black hole\footnote{We consider here black holes such as those formed in a collapse process with no second asymptotic region, so that complete spacelike slices with only one asymptotic boundary exist. } as a product of `black hole' and `exterior' Hilbert spaces $\ch_{\rm BH}$ and $\ch_{\rm ext}$, following the Minkowski decomposition into  the left and right Rindler Hilbert spaces. Roughly speaking, one expects the tensor product factorization,
\be\label{mcr} \ch_{\rm full}=\ch_{\rm ext}\otimes \ch_{\rm BH}.\ee
For full quantum gravity, or even for linearized gravitons, it is not understood how to make such a decomposition. Nevertheless, in the stringy cases if \sd\ is taken to be the outer boundary of an AdS region, a practical working knowledge of how to proceed is well-established. 

Let us nevertheless imagine  that we have achieved such a decomposition which makes sense at leading semiclassical order for any of the above choices of \sd. A state in the full Hilbert space may then be expressed as a sum over product states\footnote{Very likely we will actually need an integral over Hilbert spaces corresponding to different boundary conditions on \sd\ \cite{Donnelly:2014fua,Donnelly:2015hxa,Harlow:2015lma, Blommaert:2018rsf}  but we suppress this important point for notational brevity.}
\be |\Psi_{\rm full}\>=\sum_{A,b}c_{Ab} |\Psi^A_{\rm ext}\> |\Psi^b_{\rm BH}\>.\ee
 The existence of such a decomposition is presumed in many discussions of black hole information. Consider a set of diffeos $\z_n$, defined everywhere in the spacetime,  which all vanish near spatial infinity, but in a neighborhood of \sd\ becomes a pair of Virasoros which act nontrivially on the black hole. Since the diffeos  vanish at infinity, the associated full charges  must annihilate the full quantum state
 \be\label{red} \cq(\z_n)_{\rm full} |\Psi_{\rm full}\>=0.\ee
 On the other hand, beginning with the asymptotic surface integral expression for $\cq_{\rm full} $ and  integrating by parts we have 
 \be  \cq(\z_n)_{\rm full} =\cq(\z_n)_{\rm ext} +\cq(\z_n)_{\rm BH}.\ee
Equation \eqref{red} then becomes 
\be \label{dcrx}\sum_{A,b}c_{Ab} (\cq(\z_n)_{\rm ext}|\Psi^A_{\rm ext}\> )|\Psi^b_{\rm BH}\>=-\sum_{A,b}c_{Ab} |\Psi^A_{\rm ext}\> \cq(\z_n)_{\rm BH}|\Psi^b_{\rm BH}\>.\ee
By assumption the black hole microstates transform non-trivially under the Virasoro so neither side of the equation vanishes for all $n$. 

In the generic case, absent any extra symmetries such as supersymmetry,  we expect  $\ch_{\rm BH}$ to be composed of  Virasoro representations with highest weight $h_k$, where each $h_k$  is distinct. A black hole microstate is then uniquely determined by specifying the representation in which it lies and location therein.  In that case, \eqref{dcrx} can be satisfied only if $\ch_{\rm ext}$ 
contains all the conjugate  representations, and the constant $c_{Ab}$ are chose so that  $|\Psi_{\rm full}\>$ is a Virasoro singlet.  At first the conclusion that the exterior state should transform under the Virasoro action may seem strange. But at second thought, the exterior region has an inner boundary on which $\z_n$ necessarily acts non-trivially, so this is entirely plausible.

Given this state of affairs, it follows immediately that the specific black hole microstate in 
$\ch_{\rm BH}$ is fully determined by complete measurement of the microstate in $\ch_{\rm ext}$: it is the unique element in the conjugate representation which forms a singlet with the exterior state. Instead of \eqref{mcr} we therefore have 
\be\label{mcsr} \ch_{\rm full}=\ch_{\rm ext}.\ee
That is, factorization of the Hilbert space with the inclusion of gravity fails in the most extreme possible way: there are no independent interior black hole microstates at all! This is of course a pleasing conclusion since the independent interior microstates are at the root of the information paradox.

For supersymmetric black holes, Bogomolny bounds enforce degeneracies in the weights  $h_k$ and the argument leading to \eqref{mcsr} no longer works. Nevertheless, one may hope for a related mechanism, perhaps along the lines discussed in \cite{Schwarz:1991mv,Strominger:1996bt}  using discrete rather than continuous gauge symmetries, preventing  an unwanted  independent black hole Hilbert space. 

\section{Acknowledgements}
We are grateful to Sangmin Choi, Geoffrey Comp\`ere, Peter Galison, Monica Guica, Dan Harlow, Roy Kerr, Alex Lupsasca, Juan Maldacena,  Alex Maloney,  Suvrat Raju and Maria Rodriguez for useful conversations. This paper was supported in part by DOE DE-sc0007870, the John Templeton Foundation, the Black Hole Initiative at Harvard, the Radcliffe Institute for Advanced Study and the UK STFC.

We would also like to thank the George and Cynthia Mitchell Foundation, whose generous support allowed for over a decade of scientific workshops, spanning Cooks' Branch in Texas and Great Brampton House and Brinsop Court in England, without which this collaboration would not have been possible. The first paper in this series began at a Mitchell Foundation retreat and this current work was completed at another. We are profoundly grateful to the Foundation and personally to Sheridan Lorenz and George Mitchell for their contribution to the development of fundamental science.

\section{Appendix - Construction of covariant charges}
We begin with a very brief recap of the covariant phase space charges. A  recent comprehensive discussion, including counterterm ambiguities and also adapted to black hole horizons, can be found in \cite{Chandrasekaran:2018aop}. The starting point is the Einstein-Hilbert  Lagrangian four-form,
\bea 
L = { \e \over 16 \pi }  R.
\eea 
The  variation is 
\bea 
\delta L = - {\e \over 16 \pi }  G^{ab} h_{ab} + d\theta[h,g],
\eea 
where $\delta_h$ generates the variation $g_{ab} \to g_{ab}-h_{ab}$. The presymplectic potential $\theta$ is the three form
\bea 
\theta[h,g] = *{1 \over 16 \pi } (\n^b h_{ab}\, - \n_a h )dx^a,
\eea 
with * being the Hodge dual. The infinite-dimensional phase space of general relativity has as its tangent vectors the infinitesimal metric perturbations $h_{ab}$ that obey the linearised Einstein equations. Although $\theta$ is a three-form in
spacetime, it is also a one-form in the phase space.
The presymplectic form,
\bea 
\omega[h_1, h_2, g] = \delta_{h_1} \theta[h_2, g] - \delta_{h_2}\theta[h_1,g]
\eea 
obeys $d\omega=0$ and can therefore be used to define a conserved inner product. $\omega$ is
a two-form in the phase space.
The linearized charge $\delta \cq_0$ is then obtained from the presymplectic form 
$\omega (h_1, h_2 ;g)$ by the replacement of $h_1$ with a large diffeomorphism $\cl_\z g$,
\be \label{rf}\delta \cq_0(\z,h;g)=\int_{\Sigma_3}\omega (\cl_\z g,h;g),\ee
where we will take  $\Sigma_3$ to be a Cauchy surface for the black hole exterior with boundaries at spatial infinity and at the bifurcation surface, $\sb$. Moreover, we restrict our phase space to the on-shell perturbations $h_{ab}$ that remain finite at the boundary, up to pure gauge transformations of the form \eqref{tri}.                
When the variation is due to a diffeomorphism $\z$, the presymplectic form is exact and thus  reduces to a boundary integral, giving rise to the Iyer-Wald charge,
\be \delta \cq_{IW}(\z,h;g)={1 \over 16\pi }\int_{\p\Sigma_3}*F_{IW},\ee
as it must  in order for diffeomorphisms which vanish on the boundary to have $\delta \cq_{IW}=0$. 
Explicitly, 
\be \begin{split} F_{IWab} = \frac{1}{2}\n_a\zeta_bh
+\n_ah^c{}_b \zeta_c
+\n_c\zeta_a\ h^c{}_b 
+\n_ch^c{}_a\ \zeta_b
-\n_ah\ \zeta_b - a \leftrightarrow b. \end{split} \ee

Wald and Zoupas \cite{wz} noted an ambiguity in the addition of a possible counterterm $\delta \cq_X $  of the general form \be \delta \cq_X={1 \over 16\pi }\int_{\p\Sigma}\iota_\z( * X),\ee where $X$ is a to-be-determined spacetime one-form constructed from the geometry. The resulting charge is
\be \delta\cq = \delta \cq_{IW} + \delta \cq_X .\ee 
The interpretation of $\delta\cq$ is the difference in the charge conjugate to $\z$ between the configuration $g_{ab}$ and $g_{ab}-h_{ab}$.

  \end{document}